\documentclass[pra,twocolumn,superscriptaddress,floatfix,amsmath,amssymb]{revtex4-1}

\usepackage[english]{babel}
\usepackage{hyperref}	
\usepackage{graphicx}
\usepackage{dcolumn}
\usepackage{bm}
\usepackage{hyperref}
\usepackage{braket}
\usepackage{amsmath,amssymb,latexsym}
\usepackage{cancel}
\usepackage{bbold}
\usepackage[usenames, dvipsnames]{color}
\usepackage{amsfonts}
\usepackage{comment}
\usepackage{natbib}
\usepackage{mathrsfs}
\usepackage{soul}
\usepackage[export]{adjustbox}
\usepackage{verbatim}

\usepackage{xcolor}

\begin{document}


\title{A quantum algorithm to train neural networks using low-depth circuits}

\author{Guillaume Verdon}

 \affiliation{Institute for Quantum Computing, University of Waterloo, Waterloo, Ontario, N2L 3G1, Canada
}
 \affiliation{Department of Applied Mathematics, University of Waterloo, Waterloo, Ontario, N2L 3G1, Canada
}
 \affiliation{Perimeter Institute for Theoretical Physics, Waterloo, Ontario, N2L 2Y5, Canada
}

\author{Michael Broughton}
 \affiliation{Department of Computer Science, University of Waterloo, Waterloo, Ontario, N2L 3G1, Canada
}

\author{Jacob Biamonte}
 \affiliation{Deep Quantum Laboratory, Skolkovo Institute of Science and Technology, Moscow, 143026, Russia
}


\begin{abstract} 

Can near-term gate model based quantum processors offer quantum advantage for practical applications in the pre-fault tolerance noise regime? A class of algorithms which have shown some promise in this regard are the so-called classical-quantum hybrid variational algorithms. Here we develop a low-depth quantum algorithm to generative neural networks using variational quantum circuits. We introduce a method which employs the quantum approximate optimization algorithm as a subroutine in order produce then sample low-energy distributions of Ising Hamiltonians. We sample these states to train neural networks and demonstrate training convergence for numerically simulated noisy circuits with depolarizing errors of rates of up to $4\%$.

\end{abstract}

\maketitle
 


Multiple quantum enhanced algorithms have been proposed to speed up certain machine learning tasks \cite{2017Natur.549..195B} and existing quantum processors have now reached intermediate sizes.  The vast majority of quantum enhanced algorithms were developed for fault-tolerant quantum computing.  However, near-term quantum devices will not be error corrected and will hence inherently experience certain and often high-levels of errors during operation.  Can quantum devices with pre-fault-tolerance-threshold noise levels be useful for industrial applications or not?  

Recent classical-quantum hybrid algorithms, such as the Quantum Approximate Optimization Algorithm \cite{2014arXiv1411.4028F} and the Variational Quantum Eigensolver \cite{2016NJPh...18b3023M}, provide evidence that for some applications in optimization and quantum chemistry, there might exist a quantum advantage that is partially robust to noise. As for machine learning applications, annealers have been shown to be able to perform certain types of machine learning, but the question remained open whether or not near-term circuit model quantum computers would be able to accomplish similar tasks. This is what we demonstrate in this paper, we demonstrate how machine learning with energy-based neural network models can be performed on noisy intermediate scale quantum devices. 

Our algorithm, which we call the Quantum Approximate Boltzmann Machine (QABoM) algorithm, generates approximate samples of distributions for use in machine learning on a near-term circuit model device rather than a quantum annealer. We do so by building upon a shallow depth quantum algorithm, the Quantum Approximate Optimization Algorithm \cite{2014arXiv1411.4028F} (QAOA; which generalizes to the Quantum Alternating Operator Ansatz \cite{2017arXiv170903489H}).

QAOA can be seen as a bang-bang control variant of the quantum adiabatic algorithm \cite{2017PhRvX...7b1027Y}. It can be thought of as a coarsely Trotterized simulated adiabatic time evolution where one optimizes the pulse lengths variationally in order to approximately accomplish the same task as the adiabatic evolution with as few gates as possible \cite{2014arXiv1411.4028F}. Thus far, QAOA has been used to find the ground states of operators which encode problem instances---in this paper we extend its capabilities to generating a quantum distribution which can be sampled in machine learning.   An interesting property of QAOA is that one might find optima of cost functions in a way where the classical optimization overhead is near-constant, by e.g.~keeping the number of pulses and the optimization run-time fixed.

{\it Structure.} We will now present background material on Quantum Boltzman Machines, as they have been used in quantum annealing.  Then we will explain the basic building blocks of Quantum Approximate Optimization.  In Section \ref{sec:algos} the main algorithms are presented, followed by a section devoted to presenting the results of our numerical experiments. A discussion section precedes the conclusion which is followed by an appendix providing additional details that can aid in reproducing the results of this study.  All source code and data is available open-source and linked to in the references.   

\section{Background}
\subsection{Quantum Boltzmann Machines}

Boltzmann machines are a tyae of generative neural network learning model in which an interacting collection of spins---representing bits of data---are typically trained to associate a low-energy to the spin-representations of a training data-set distribution.  After thermalization---a process thought to be accelerated by quantum computers \cite{2016arXiv160905537B}---a Boltzmann machine can be sampled to produce and also to recognize patterns.  Training a network of quantum spins such that this network of spins will assign low-energy values to an entire training set is precisely the computational bottleneck of Boltzmann-machine-based deep learning.  

The approaches to train a Boltzmann machine rely on sampling the distribution which is thermal with respect to the network's energy function. From this procedure, a model which approximates the data and its correlation structure is obtained. The energy or Hamiltonian function---given as a linear matrix representation in the quantum case---is often chosen to be that of an Ising model, i.e.~a symmetric Hamiltonian diagonal in the standard basis and of the form 
\begin{equation}\label{Ising}
    \hat{H}\equiv-\sum_{j,k\in u} J_{jk} \hat{Z}_j\hat{Z}_k - \sum_{j\in u} B_j \hat{Z}_j
\end{equation}
where $u$ is an index set for the vertices of a neural network graph $\mathcal{G}$ and $\hat{Z}$ is a Pauli-Z operator. The subset of spins representing data are called \textit{visible} units, while all the rest are called \textit{hidden} units. 
Mathematically, the goal of the Boltzmann machine algorithm (quantum as well as classical) is for the reduced thermal state on the visibles $\rho|_v = \text{tr}_h \left(e^{-\beta H}\right)/\text{tr}(e^{-\beta H})$ to approximate the state representing the normalized sum over all the data $\rho_{\text{data}}= |D|^{-1}\sum_{d_j\in D} \ket{d_j}\bra{d_j}$, where the  non-empty indexed data set was labelled as $D = \{d_j\}_j$. To quantify the statistical distance between the visibles' reduced state and the training data, we can use the quantum relative entropy, the quantum analogue of the Kullback-Leibler divergence. By sampling the Gibbs state of the Ising Hamiltonian for a given choice of parameters $\bm{\theta} = \{J_{jk}, B_j\}_{j,k\in u}$, one can then compare the statistics of the reduced state on the visible units to that of the data and then suggest weight updates to reduce the relative entropy (and hence train the network).  

\begin{figure}[h!]
 \begin{center}
\includegraphics[width=0.75
\columnwidth]{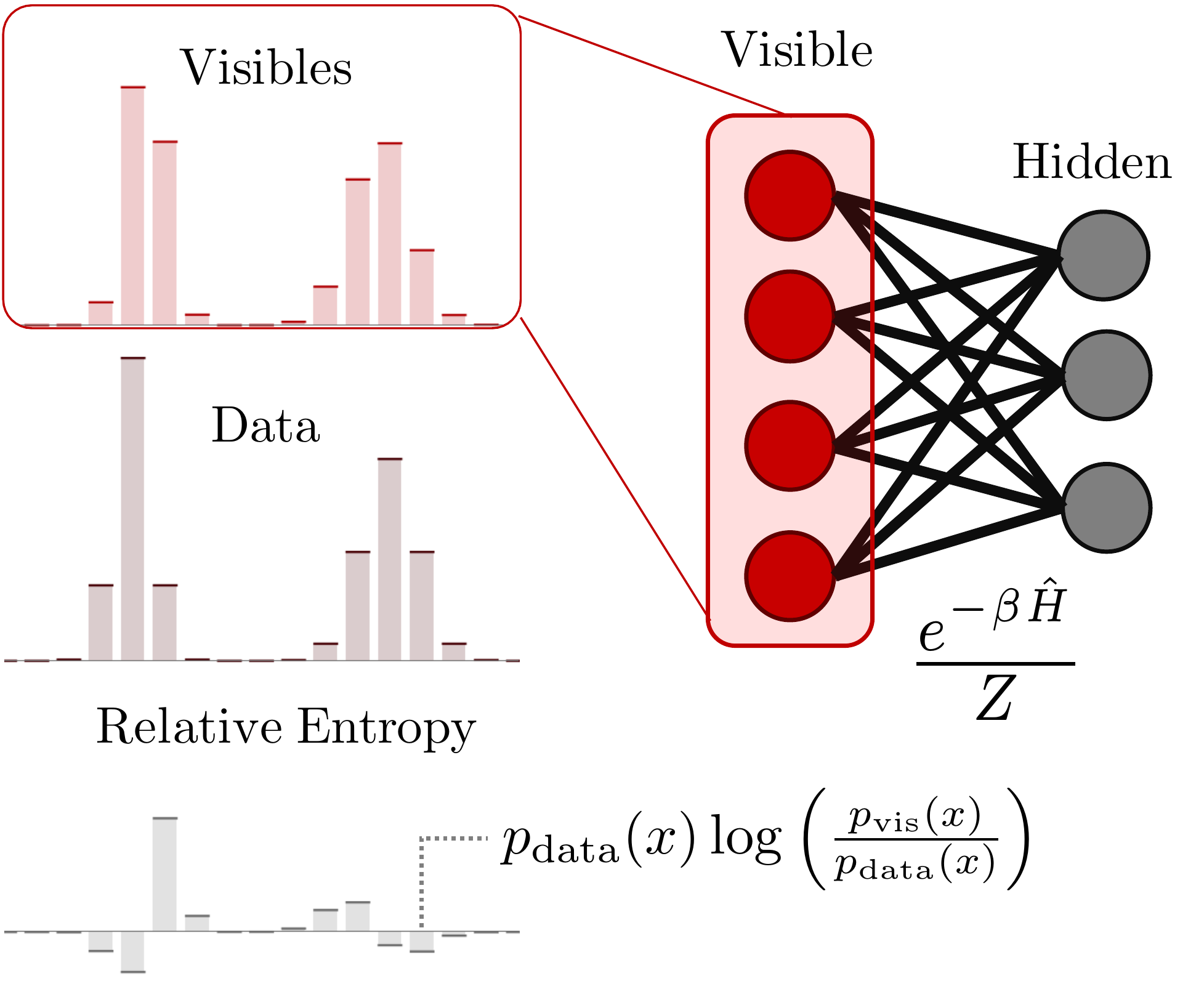}
\caption{Representation of a Restricted Boltzmann Machine (RBM) neural network, a distribution over states of the visible units, a data distribution, and the relative entropy density between these distributions. An RBM is a Boltzmann machine with a complete bipartite graph topology. The relative entropy between the data and the marginal Gibbs state on the visible units is the measure of statistical distance to be minimized for effective learning.
} \label{fig:sampling}
\end{center}
\end{figure}
 
Instead of minimizing the relative entropy between the visibles and the data by computing derivatives of the relative entropy, it is rather simpler to minimize a lower bound on this relative entropy \cite{QRBM}.  This method consists of updating each component of $\bm{\theta}$ by comparing expectation values of $\langle \partial_{\bm{\theta}} \hat{H}\rangle$ for the thermal state of $\hat{H}$, to that of the average statistics of a thermal state with so-called \textit{clamped} visibles. 

The procedures consisting of collecting datapoints to estimate these expectation values are called \textit{unclamped} and \textit{clamped} Gibbs sampling, respectively. Clamped sampling can be achieved by simulating thermalization with respect to $\hat{H}$ with an added \textit{clamping potential}, $\hat{V}_j = -\log \ket{d_j}\!\bra{d_j}_v$ for each data point $d_j\in D$. The update rule for each of the parameters is $\theta_k \mapsto \theta_k+\delta\theta_k$, where 
\[\delta \theta_k = \tfrac{1}{|D|}\sum_{x\in D}\braket{\partial_{\theta_k} \hat{H} }_{\hat{H}+\hat{V}_x} -\braket{\partial_{\theta_k} \hat{H} }_{\hat{H}}.\]
The above expectation(s) are taken for the state which is thermal according to the subscript Hamiltonian, i.e.~$\braket{\ldots}_{\hat{K}} \equiv \text{tr}(e^{-\beta \hat{K}}\ldots)/\text{tr}(e^{-\beta \hat{K}})$. Estimating these expectation values of various observables with respect to Gibbs states is where quantum computers can be used to accelerate the training. We will describe how quantum circuits can be used to achieve this Gibbs sampling approximately in Section \ref{sec:algos}.

\subsection{Quantum Approximate Optimization}

The Quantum Approximate Optimization Algorithm was originally proposed to solve instances of the MaxCut problem. The QAOA framework has since been extended to encompass multiple problem classes related to finding the low-energy states of Ising Hamiltonians. 

In general, the goal of the algorithm is to find approximate minima of a pseudo Boolean function $f$ on $n$ bits, $f(\bm{z})$, $\bm{z}\in\{-1,1\}^{\times n}$. This  function is often an $m^{\text{th}}$-order polynomial of  binary variables for some positive integer $m$, e.g., $f(\bm{z}) = \sum_{p\in\{0,1\}^m}\alpha_{\bm{p}}\bm{z}^{\bm{p}}$, where $\bm{z}^{\bm{p}}=\prod_{j=1}^n z_j^{p_j}$. The case where this polynomial is quadratic ($m=2$) has been extensively explored in the literature, and will be the main focus in this paper. 

The QAOA approache to optimization first starts in an initial product state $\ket{\psi_0}^{\otimes n}$ and then a tunable gate sequence produce a wavefunction with a high probability of being measured in a low-energy state (with respect to a cost Hamiltonian). We define the energy to be minimized as the expectation value of the cost Hamiltonian $\hat{H}_C \equiv f(\bm{\hat{\bm{Z}}})$, where $\bm{\hat{Z}} = \{\hat{Z}_j\}_{j=1}^n$. 

\begin{figure}[h!]
\includegraphics[width=0.65\columnwidth]{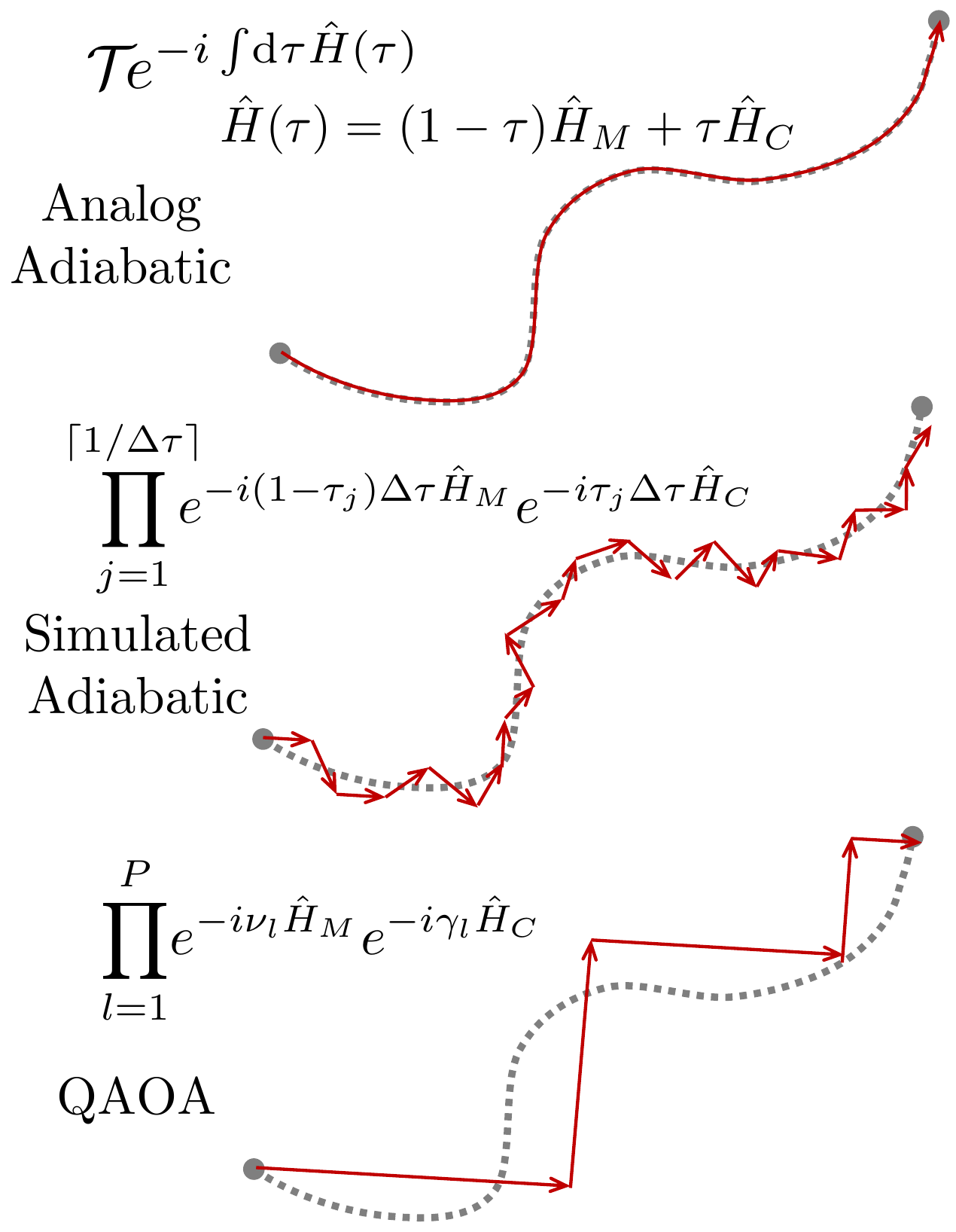}
\begin{center}
\caption{Conceptual analogy for comparing (bottom) QAOA to (top) analog adiabatic and (middle) quantum simulated adiabatic evolution as a path through state space. 
} \label{fig:sampling}
\end{center}
\end{figure}

\section{Algorithms}\label{sec:algos}
\subsection{Quantum Variational Thermalization}

The goal of quantum variational thermalization is to variationally approximate the statistics of the thermal state of a given Hamiltonian. In our case, we would like to approximate the statistics of the thermal state $\hat{\rho}_\beta = e^{-\beta \hat{H}_C}/\text{tr}(e^{-\beta \hat{H}_C})$ of the Ising Hamiltonian ~$\hat{H}_C\equiv-\sum_{j,k} J_{jk} \hat{Z}_j\hat{Z}_k - \sum_{j} B_j \hat{Z}_j$. Our approach will consist of optimizing over a family of states $\hat{\rho}(\bm{\mu})$ where $\bm{\mu}$ is a set of variational parameters for our state preparation ansatz. We will use the free energy $F_\beta$ as our error function to be minimized through the variation of parameters. Note that the free energy at inverse temperature $\beta$ is proportional to the temperature times the relative entropy \footnote{We use the natural logarithm in our relative entropy for convenience. Hence, our units of information are in nats rather than bits.} of the state relative to to the thermal state, $F_\beta(\hat{\rho}) = \tfrac{1}{\beta}D_{K\!L}(\hat{\rho}|| \hat{\rho}_\beta)$, hence minimizing the free energy at fixed temperature will minimize the relative entropy to the thermal state. Note that our interest lies in the classical statistics of the samples in the computational basis, hence our free energy is the \textit{Shannon} free energy, namely $F_\beta(\hat{\rho})  = \braket{\hat{H}_C}-\tfrac{1}{\beta}S$.

In order to variationally minimize the free energy, we propose two approaches. One keeps the Von Neumann entropy of the system fixed while the energy is minimized with a QAOA loop. The second method variationally adapts the input Von Neumann entropy. 

Our algorithm consists of first preparing an exact thermal state of an initial uncoupled Hamiltonian $\hat{H}_I$, e.g.~$\hat{H}_I = \sum_j \hat{Z}_j$, then using QAOA to minimize the energy expectation with respect to a \textit{cost Hamiltonian} $\hat{H}_C$, e.g.~$\hat{H}_C\equiv-\sum_{j,k} J_{jk} \hat{Z}_j\hat{Z}_k - \sum_{j} B_j \hat{Z}_j$, which consists of applying alternating exponentials of the cost and of a \textit{mixer} Hamiltonian, e.g., $\hat{H}_M = \sum_j \hat{X}_j$. This allows the sampling of a state which \textit{approximates} the thermal state statistics of the cost Hamiltonian $e^{-\beta \hat{H}_C}/\text{tr}(e^{-\beta \hat{H}_C})$, when measured in the standard basis.

For the variational energy fixed entropy variant, explicitly, we start by preparing the state $
 \ket{\psi_0}=\bigotimes_{j} \sqrt{2\cosh(\beta)} \textstyle\sum_{b\in\{0,1\}} e^{(-1)^{1+b} \beta/2 }\ket{b}_j\ket{b}_{E_j}$ 
using a set of environment purification registers $E = \bigotimes_{j} E_j$ of equal number of qubits to that of the system to be thermalized. This state is efficient to prepare as it is of low constant depth. Tracing this state over the environment qubits, one recovers the thermal state of $\hat{H}_I$, i.e. $e^{-\beta \hat{H}_I}/\text{tr}(e^{-\beta \hat{H}_I})$. Which is equal to $\bigotimes_{j} \frac{1}{2}\sum_{b\in\{0,1\}}\left(1-(-1)^b \tanh (\beta)\right) \ket{b}_j\!\bra{b}_j$.

Following this initial thermal state preparation, we apply the QAOA to minimize the expectation value $\braket{\hat{H}_C}$, with $\hat{H}_M$ as the mixer Hamiltonian and $\hat{H}_C$ as the cost Hamiltonian. This consists of applying the operations
$\prod_{l=1}^P \exp(-i\nu_l \hat{H}_M)\exp(-i\gamma_l \hat{H}_C)$ for some fixed $P$ and some real parameters $\{\bm{\nu},\bm{\gamma}\}$, then measuring the system in the computational basis, repeating this preparation and measurement $N$ times to get an estimate of $\braket{\hat{H}_C}$, and using $M$ steps of a classical optimization algorithm (such as Nelder-Mead \cite{Nelder_1965}) to vary the parameters $\{\bm{\nu},\bm{\gamma}\}$ in a way to minimize $\braket{\hat{H}_C}$.

After this QAOA, we claim that the final state for this set of pulses (even crudely) approximates the thermal state of $\hat{H}_C$, in the sense that measuring local observables of this state yields expectation values approximating that of the state $e^{-\beta \hat{H}_C}/\text{tr}(e^{-\beta \hat{H}_C})$, to an accuracy that is sufficient for sampling in the context of neural network training. We demonstrate this empirically through our numerical experiments. In the appendix we explore in greater detail what the final state after QAOA optimization would look like in the asymptotic large-depth QAOA case $P\rightarrow \infty$. It is believed that in this regime the QAOA behaves effectively \cite{2014arXiv1411.4028F} like a gapped adiabatic evolution under an interpolating Hamiltonian. We compare this to analog simulated thermalization, which is the evolution performed by physical quantum annealers (such as D-Wave devices \cite{2016arXiv161104528K}), which can also be used for Gibbs sampling. 

\subsection{Quantum Approximate Boltzmann Machine} We now describe the main algorithm.  There are two variants of the algorithm, (i) a gate-based analogy of the Quantum Boltzmann Machine algorithm \cite{QRBM}, and (ii) a \textit{quantum randomized clamping} (QRC) variant of the same algorithm, where the training is performed with batches of data at a time, and the input is randomized either through classical randomization or through the use of a Quantum Random Access Memory \cite{2008PhRvL.100p0501G}. 

Let $v$ and $h$ be the sets of indices for the visible and hidden units. Let $u = v\cup h$ be the set of indices for all units. Let $D$ be the dataset made of bit strings $\bm{d}\in D$ of length number of visible units $|v|$. 

In the first variant (i), we begin by initializing the network parameters (weights and biases) randomly, providing the zeroth epoch parameters $\bm{\theta}^{(0)}$. Alternatively, one might perform a grid search over random weights which provides a better loss, or perform any other form of hyperparameter optimization, standard in machine learning \cite{LeCun_2015,Hinton_2012,Goodfellow-et-al-2016}. Each weight update depends on computing expectation values of certain terms in the cost Hamiltonian of a given epoch. The averages to be computed are for equilibrium with respect to the \textit{clamped} and \textit{unclamped} averages. Both sampling procedures are done via our QAOA-based quantum approximate thermalization, in each case a different cost and mixer Hamiltonian is used.

At each epoch, we have a set of network parameters $\bm{\theta}^{(n)}$. Given these network parameters, we can define the \textit{full} and \textit{partial} cost Hamiltonians for epoch $n$, the full Hamiltonian is 
\begin{equation}
\hat{H}^{(n)}_C\equiv-\sum_{j,k\in u} J^{(n)}_{jk} \hat{Z}_j\hat{Z}_k - \sum_{j\in u} B^{(n)}_j \hat{Z}_j 
\end{equation}
while the partial cost Hamiltonian $\hat{H}_{\tilde{C}}^{(n)}$ excludes terms strictly supported on the visibles. See appendix \ref{app:reg} for the full form of this cost Hamiltonian. These Hamiltonians are used to perform QAOA for the unclamped and clamped sampling respectively. The QAOA mixer Hamiltonians, for the unclamped and clamped Gibbs sampling, which we call the full and partial mixer Hamiltonians, are given by  $\hat{H}_M = \sum_{j\in u} \hat{X}_j$ and $\hat{H}_{\tilde{M}} = \sum_{j\in h} \hat{X}_j$ respectively. Again the partial Hamiltonian is like the full Hamiltonian with the terms on the visibles removed. The algorithms for the clamped and unclamped sampling are closely related; as they both rely on a QAOA subroutine with similar Hamiltonians. 

We begin by describing the process of unclamped sampling.
First, using a set of $|u|$ ancillary qubits, by creating some partially entangled Bell pairs, we prepare the thermal state $\hat{\rho}_I = \mathcal{Z}^{-1}_Ie^{-\beta \hat{H}_I} =\bigotimes_{j} \mathcal{Z}_j^{-1}e^{-\beta \hat{Z}_j}$ where $\mathcal{Z}_I= \prod_j \mathcal{Z}_j,\ \mathcal{Z}_j = \text{sech}(\beta)/2$. Following this, for each epoch $n$, we apply the QAOA algorithm with our full cost and full mixer Hamiltonians, the $m^\text{th}$ QAOA iteration of the $n^\text{th}$ epoch consists of applying the pulses
\begin{equation}\hat{U}^{(n,m)}_{\bm{\nu},\bm{\gamma}} \equiv \prod_{l=1}^P \exp(-i\nu^{(n,m)}_l \hat{H}_M)\exp(-i\gamma^{(n,m)}_l \hat{H}_C).\end{equation}
After the pulses are applied, we measure the cost Hamiltonian expectation value
\begin{equation}
    \braket{\hat{H}^{(n)}_C}_{(n,m)} = \text{tr}(\hat{U}^{(n,m)\dagger}_{\bm{\nu},\bm{\gamma}}\hat{H}_C \hat{U}^{(n,m)}_{\bm{\nu},\bm{\gamma}}\hat{\rho}_M)
\end{equation}
via the expectation estimation algorithm (QEE) \cite{2016NJPh...18b3023M}. The QEE algorithm consists of estimating expectation values of individual terms in the Hamiltonian via repeated identical state preparations followed by measurements, and classically summing up these up to get an estimate for the global expectation value.
 The pulse parameters are updated using a classical optimizer, such as Nelder-Mead \cite{Nelder_1965}, to minimize $\braket{\hat{H}^{(n)}_C}_{(n,m)}$, for a number of optimization iterations $M$. We then repeat the state preparation and measurement with these new parameters $\bm{\gamma}^{(n,m+1)}$ and $\bm{\nu}^{(n,m+1)}$.
The first set of pulse parameters for a given weight epoch $n$, i.e., $\bm{\gamma}^{(n,0)}$ and $\bm{\nu}^{(n,0)}$ are initialized as random.
Once an optimum of  $\braket{\hat{H}^{(n)}_C}$ is deemed reached; the optimal $\bm{\gamma}^{(n)}$ and $\bm{\nu}^{(n)}$ QAOA parameters have been found for epoch $n$. At this point we have the full circuit to perform Gibbs sampling for our weight updates. We can then measure the unclamped expectation values $\braket{\hat{Z}_j\hat{Z}_k}_{(n)}$ and $\braket{\hat{Z}_j}_{(n)}$ for this optimal QAOA circuit for epoch $n$. Thus we have explained how to perform \textit{unclamped} Gibbs sampling.

For \textit{clamped} Gibbs sampling, the algorithm differs in every case where the mixer Hamiltonian and the cost Hamiltonian were used: they  are replaced with the partial mixer and partial cost Hamiltonians. To sample the Gibbs distribution of the clamped Hamiltonian for data point $x\in D$, we initially prepare a thermal state of the hidden units $\sim e^{-\beta H_{\tilde{I}}}$, $H_{\tilde{I}}\equiv \sum_{j\in h} Z_j$ via partially entangled Bell pairs, which leaves the hidden units in a mixed state, while preparing the visible units in the computational basis state $\ket{x}_v$. The same QAOA routine as the unclamped sampling is applied, except with the partial mixer and cost Hamiltonians $H_{\tilde{M}}$, and $H_{\tilde{C}}$. At a given epoch $n$ we can sample the expectation values for the optimal partial cost minimizing QAOA pulse sequence, $\braket{Z_jZ_k}_{(n),x}$ and $\braket{Z_j}_{(n),x}$. We repeat this QAOA optimization and sampling for each data point.

Once the expectation values for the unclamped case and the clamped case for each data point is estimated, we can then update the weights according to Melko et al.'s \cite{QRBM} bound-based QBM rule., i.e. 
\begin{align}
\delta J^{(n)}_{jk} 
&= \overline{\braket{Z_j Z_k}}_D - \braket{Z_j Z_k}\\
  \delta B^{(n)}_{j} &= \overline{\braket{Z_j }}_D - \braket{Z_j}
\end{align}
and the $(n+1)^\text{th}$ epoch's weights are then   $J^{(n+1)}_{j} = J^{(n)}_{j}+ \delta J^{(n)}_{j}$, and
$B^{(n+1)}_{j} = B^{(n)}_{j}+ \delta B^{(n)}_{j}$.

The regular training algorithm can be parallelized over multiple quantum chips aided by classical computers, each running QAOA optimization for each data point to compute each gradient update step. The clamping of each data point is done in a simulated fashion by preparing the initial state of the visible units in the $\ket{\bm{x}}$ state (step 3 (b)). Instead of clamping to a single data point at a time, we can perform \textit{Quantum Randomized Clamping} (QRC), this allows us to train all data points (or a randomly chosen subset; also known as a \textit{minibatch}) with one QAOA optimization. 

One option for this Quantum Randomized Clamping (QRC) is to use a Quantum Random Access Memory \cite{2008PhRvL.100p0501G}.
For a dataset $D=\{d_j\}_j$, using a QRAM, in a $\mathcal{O}(\log |D|)$ gate depth, we can prepare a state $|D|^{-1/2}\sum_{j=0}^{D-1} \ket{j}_A\ket{d_j}_V$ where $A$ is a binary address register. We can feed the $V$ register to the visible units, and run the rest of the algorithm similarly, except that the averaging of expectation values over all data points will be done automatically. 

Another option to the same effect is to classically randomly pick a certain data point $d_j$ to clamp our visibles to, for each measurement iteration of the QEE, for each QAOA update, for each weight update. This effectively is akin to preparing the state $|D|^{-1}\sum_{d_j\in D} \ket{d_j}\! \bra{d_j}_V$ and simulating thermalization with this mixed state clamped for the visibles.

Since we have to run QAOA only twice for each gradient update in version (ii) as compared to version (i) needing $1+|D|$ different QAOA optimizations, this can be seen as a speedup over the traditional clamping algorithm, albeit at perhaps a cost of greater difficulty of QAOA optimizations. In appendix \ref{app:QRC} we examine in greater detail the relation between both of our  approaches to randomized clamping.

\section{Numerical Experiments}
\begin{figure}[h!]
 \begin{center}
\includegraphics[width=1.02\columnwidth]{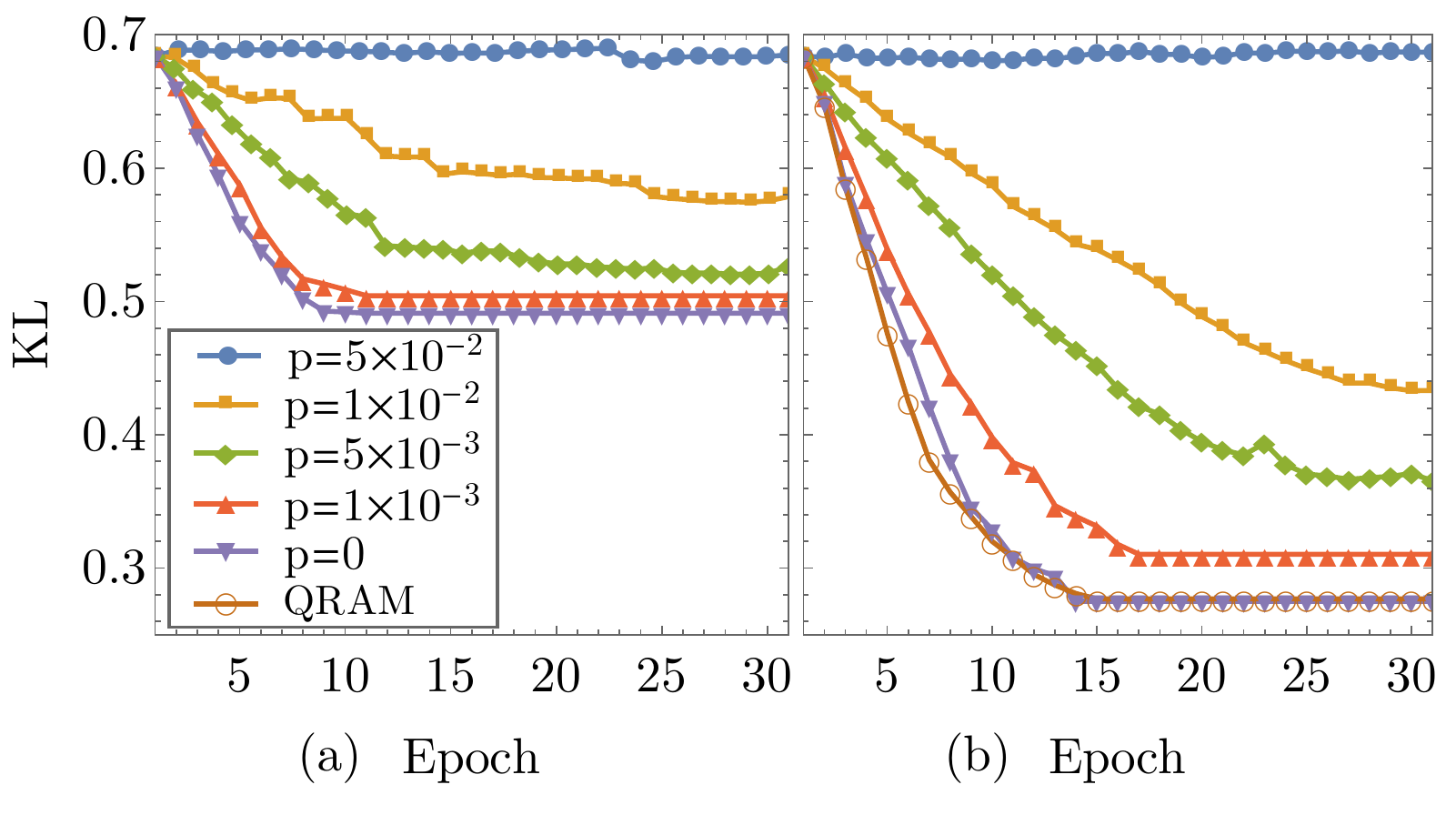}
\caption{Kullback-Leibler divergence of reconstructed distribution relative to data, versus training epoch, for various levels of depolarizing noise ($\mathcal{N}_p$) (a) for regular clamping training (b) for quantum randomized clamping (QRC), including a noiseless case with QRAM-aided QRC. Both plots share the same vertical scale. 
} \label{fig:KL_noises}
\end{center}
\end{figure}

Figure \ref{fig:KL_noises} depicts training a restricted Boltzmann machine with both variants of the QABoM algorithm. The Kullback-Leibler (KL) divergence is computed by performing inference classically using standard techniques of Restricted Boltzmann Machines \cite{Hinton_2012}, using the weights trained on the quantum computer at each epoch. Note that for the specific case of restricted Boltzmann machines, the clamped sampling can be done efficiently classically, but we opted to perform it using our algorithm, as this would be needed for more general network topologies, such as semi-restricted or full/deep Boltzmann machines \cite{QRBM, LeCun_2015}.

The number of measurements per QAOA update was $N=500$, with QAOA depth $P=3$, the number of Nelder-Mead optimizations per QAOA parameter update was $M=100$. The circuit was compiled with a probability $p$ of applying each Pauli Error, i.e.~each gate has the depolarizing channel added 
\begin{equation}
\mathcal{N}_p(\rho) = (1-3p)\rho+pX\rho X+pY\rho Y +pZ\rho Z.\end{equation} An alternate way to write this channel is $\mathcal{N}_p(\rho) = (1-4p)\rho +4p (I/2)$, this gives an average gate fidelity $\bar{F}_1 = (1-2p)$ and $\bar{F}_2=(1-2p)^2$ for 1 and 2-qubit gates respectively. All cases with $p\leq 1\%$ showed signs of training convergence. In some cases the training updates were terminated once the minimal value of KL divergence was reached, as tested with new data points. The network consisted of 4 visible units and 2 hidden units. We see that the version of the training with QRC outperforms the regular training algorithm. This shows that the randomized clamping provides weight updates that better approximate the KL gradient as compared to the regular \cite{QRBM} bound-based update rule using single-data-point clamping.

 Figure \ref{fig:sampling} examines the scaling of the quality of the Quantum Expectation estimation with increasing measurements (right), and if we try to scale up the number of QAOA pulses (left), with the number of Nelder-Mead iterations fixed to $M=100$, for various noise levels. For the Quantum Expectation Estimation, we depict the average error in weight update, measured in the squared Euclidean $\mathbb{R}^T$ norm, where $T=\text{dim}(\bm{\theta})$.
 
We see that in the noisy case the extra depth is detrimental, while in the noiseless case due to increased optimization difficulty and fixed $M$ we get a slightly worse performance. This could be perhaps partially medied by the use of a different optimization algorithm than Nelder-Mead.

\begin{figure}[h!]
 \begin{center}
\includegraphics[width=1.02\columnwidth]{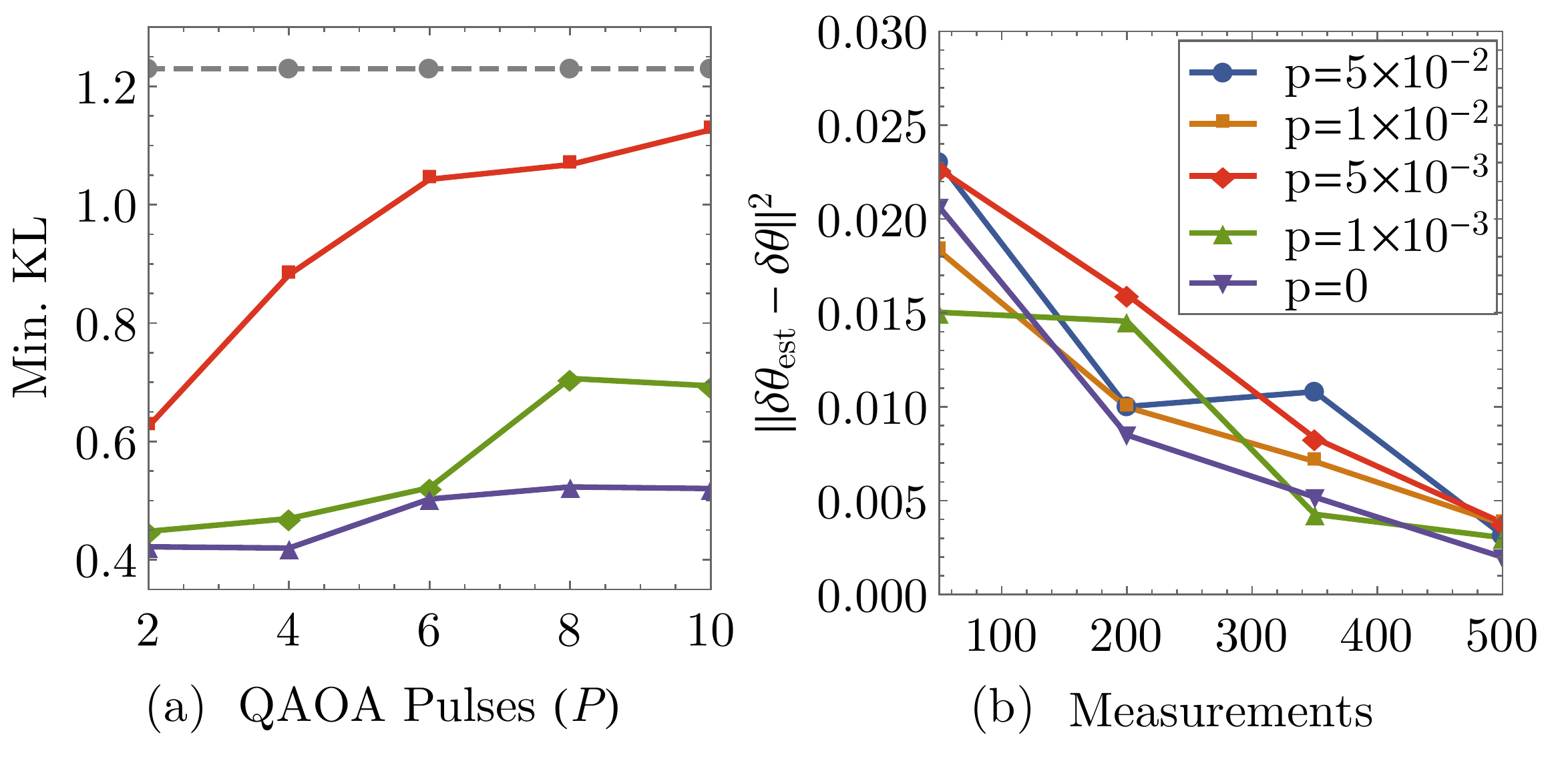}
\caption{For various depolarizing noise levels ($\mathcal{N}_p$), training with hidden mode data \cite{QRBM}: (a) Minimum KL achieved versus number of QAOA pulses. Dotted line is initial KL. Number of Nelder-Mead steps fixed to $M=100$. (b) Euclidean norm squared of error in weight update due to quantum expectation estimation versus number of measurements.  $\delta\theta_\text{est}$ is the weight update calculated from these estimated expectation values through measurements, while $\delta\theta$ is the weight update for the actual expectation values calculated directly from the simulated wavefunctions. 
} \label{fig:sampling}
\end{center}
\end{figure}

\section{Discussion} Sampling exact thermal states of quantum systems, such as if one were to use the quantum metropolis algorithm \cite{2011Natur.471...87T}, is still futuristic. For near-term quantum computing devices, our algorithm provides a means to train neural networks using noisy devices in an approximate way.  This achieves practical levels of learning, as we have demonstrated through numerical simulation of noisy quantum computation.

Our algorithm prepared crude pseudo-thermal states, not exact thermal states, yet this achieved levels of learning performance. A possible extension would be to attribute our near-thermal performance to the Eigenstate Thermalization Hypothesis \cite{2017arXiv171004631B}, we leave this connection to be fleshed out in future work. An extension of this algorithm to improve proximity to thermality of our states could be to variationally maximize the entropy at fixed energy, assuming one could get estimates of the entropy efficiently.

We focused on a restricted Boltzmann machines. Our QABoM algorithm is technically general enough to allow for any sort of  training of quantum Boltzmann machine, as in \cite{QRBM}, i.e.~supervised, unsupervised, deep, restricted and semi-restricted. Thus, an extension of this work could be to test semi-restricted Boltzmann machines and other network architectures. We chose the restricted case in order to perform the inference classically and as an initial stepping stone to network architechtures of higher complexity.

Given that we demonstrated a certain robustness of our training algorithm under levels of noise comparable to that of current-term devices \cite{ 2015arXiv151004375G,reagor2018demonstration}, it is quite feasible that this algorithm be implemented in the near-future. An immediate extension to this work preceding implementation could be to test robustness under different types gate and measurement noise models (beyond depolarizing) which better reflect the observed noises of an implementation of interest.

\section{Acknowledgements} The authors would like to thank the Creative Destruction Lab for hosting the Quantum Machine Learning program, where the idea for this paper was conceived. All circuits in this paper were simulated on the Rigetti Forest API Quantum Virtual Machine and written in PyQuil \cite{2016arXiv160803355S}. We thank Will Zeng and the Rigetti Computing team for providing technical support and computing facilities to run the quantum circuit simulations.  We also thank Mohammid Amin, Jason Pye, Nathan Wiebe, Peter Johnson, Jonathan Romero, and Jonathan Olson for useful discussions. GV acknowledges financial support from NSERC. This research was supported in part by Perimeter Institute for Theoretical Physics. Research at Perimeter Institute is supported by the Government of Canada through the Department of Innovation, Science and Economic Development and by the Province of Ontario through the Ministry of Research and Innovation.

\appendix 

\section{Explicit Algorithms} In this section we describe the steps of the QABoM algorithm in full detail.
Let $v$ and $h$ be the indices for the visible and hidden units. Let $u = b\cup h$ be the set of indices for all units. Let $D$ be the dataset made of bit strings $\bm{x}\in D$ of length number of visible units $|v|$.  \\

\subsection{Regular QABoM}\label{app:reg}

\begin{itemize}
\item[-] Define the full and partial initial Hamiltonians, respectively  $H_I = \sum_{j\in u} Z_j$, and $H_{\tilde{I}}=\sum_{j\in h} Z_j$

\item[-] Define the full and partial mixer Hamiltonians, respectively $H_M = \sum_{j\in u} X_j$ and $H_{\tilde{M}} = \sum_{j\in h} X_j$

\item[-] Define epoch 0 weights $J^{(0)}_{jk}$, biases $B^{(0)}_j$ as random (or from classical pretraining) 
\item[-] For epochs $n\geq 1$, run steps 1-5 in a loop:
\end{itemize}
\begin{enumerate}
\item (Epoch $n$) Weights $J^{(n)}_{jk}$, biases $B^{(n)}_j$ given from previous epoch
\begin{enumerate}
\item 
Define the full cost Hamiltonian\[\hat{H}^{(n)}_C\equiv\sum_{j,k\in u} J^{(n)}_{jk} \hat{Z}_j\hat{Z}_k + \sum_{j\in u} B^{(n)}_j \hat{Z}_j \]

\item 
Define the partial cost Hamiltonian\[H^{(n)}_{\tilde{C}}\equiv\sum_{j,k\in u} J^{(n)}_{jk} \hat{Z}_j\hat{Z}_k + \sum_{j\in h} B^{(n)}_j \hat{Z}_j \]

\end{enumerate}

\item Unclamped thermalization
\begin{enumerate}
\item Initialize pulse parameters $\bm{\gamma}^{(n,0)}$ and $\bm{\nu}^{(n,0)}$ as random
\item Prepare a thermal state of $H_I$ of inverse temp $\beta$ by entangling qubits in pairs
\item Apply QAOA circuit via $P$ simulated pulses alternating between cost and partial  Hamiltonian evolution  
\[\qquad \qquad\prod_{l=1}^P \exp(-i\nu^{(n,m)}_l H_M)\exp(-i\gamma^{(n,m)}_l H_C)\]
\item Measure cost expectation value $\braket{\psi_{n,m}|H^{(n)}_C|\psi_{n,m}}$ via expectation estimation algorithm 
\item Using classical optimizer, figure out updates $\bm{\gamma}^{(n,m+1)}$ and $\bm{\nu}^{(n,m+1)}$ to minimize $\braket{H^{(n)}_F}$

\item Repeat (b)-(e) until minimum of $\braket{H^{(n)}_F}$ reached; optimal  $\bm{\gamma}^{(n)}$ and $\bm{\nu}^{(n)}$ QAOA parameters are found.
\item Measure \& register expectation values $\braket{Z_jZ_k}$ and $\braket{Z_j}$ for optimal QAOA circuit from (f)
\end{enumerate}

\item Clamped thermalization 

For each data string $\bm{x}\in D$:\\

\begin{enumerate}
\item Initialize pulse parameters $\bm{\gamma}^{(n,0)}$ and $\bm{\nu}^{(n,0)}$ as random
\item Prepare visible units as computational basis state corresponding to data point $\ket{\bm{x}}_v$
\item Prepare hidden units in a thermal state of $H_{\tilde{I}}$ of inverse temp $\beta$ by entangling qubits in pairs
\item Apply QAOA circuit via $P$ simulated pulses alternating between partial cost and partial mixer Hamiltonian evolution \[\qquad\qquad \prod_{l=1}^P \exp(-i\nu^{(n,m)}_l H_M)\exp(-i\gamma^{(n,m)}_l H_C)\]
\item Measure cost expectation value $\braket{\psi_{n,m}|H^{(n)}_{\tilde{C}}|\psi_{n,m}}$ via expectation estimation algorithm 
\item Using classical optimizer, figure out updates $\bm{\gamma}^{(n,m+1)}$ and $\bm{\nu}^{(n,m+1)}$ to minimize $\braket{H^{(n)}_F}$

\item Repeat (b)-(e) until minimum of $\braket{H^{(n)}_F}$ reached; optimal  $\bm{\gamma}^{(n)}$ and $\bm{\nu}^{(n)}$ QAOA parameters are found.
\item Measure \& register expectation values $\braket{Z_jZ_k}$ and $\braket{Z_j}$ for optimal QAOA circuit from (g)
\end{enumerate}

\item Update weights using Bound-based QBM rule \cite{QRBM}. Let $\braket{\ldots}$ denote the unclamped thermalization expectation values (computed in 2(g)). Let $\overline{\braket{\ldots}}_D$ denote the average over all expectations for each data point $\bm{x}\in D$ for the clamped thermalizations (computed in 3(h)). Update weights for next epoch by defining
\begin{enumerate}

\item $\delta J^{(n)}_{jk} = \overline{\braket{Z_j Z_k}}_D - \braket{Z_j Z_k}$\\
\item  $\delta B^{(n)}_{j} = \overline{\braket{Z_j }}_D - \braket{Z_j}$\\

\item  $J^{(n+1)}_{j} = J^{(n)}_{j}+ \delta J^{(n)}_{j} $\\
\item  $B^{(n+1)}_{j} = B^{(n)}_{j}+ \delta B^{(n)}_{j} $
\end{enumerate}
\item Define epoch = $n+1$, go back to step 1.
\end{enumerate}

\subsection{QRC QABoM}\label{app:QRC}

\subsubsection{Quantum Random Access Memory Randomization}

The QRAM-accelerated version of the QABoM algorithm allows for training of the whole dataset in a single run of QAOA optimization 
For a dataset $D=\{d_j\}_j$, for step 3(b), using a QRAM, in a $\mathcal{O}(\log |D|)$ gate depth, we can prepare a state \[\tfrac{1}{\sqrt{|D|}}\sum_{j=0}^{|D|-1} \ket{j}_A\ket{d_j}_V\] where $A$ is a binary address register mad eof ancillary qubits. We can feed the $V$ register to the visible units, and run the rest of the algorithm similarly, except that the averaging of expectation values over all data points ($\overline{\braket{\ldots}}_D$  step 4) will be done automatically by estimating the expectation value with the data clamped via the QRAM. 

\subsubsection{Classically Randomized Clamping}
Note that we technically do not need a QRAM to perform this version of the algorithmm. Since the QAOA with partial cost and mixer Hamiltonians leaves the visible layer untouched, and that we measure all units (visible and hidden) in the standard basis at the end of each run, the entangled state between the system and the visible layer qubits will give a certain outcome $\ket{d_j}\!\bra{d_j}_v$. Note the address register is unused in the computation, hence we can effectively trace our this register, what we are left with is the mixed state \[\tfrac{1}{|D|} \sum_{j=0}^{D-1} \ket{d_j}\!\bra{d_j}_v
\]
which is a classical mixture of our data points. Note that the QRAM-aided clamping is effectively like the purification of the classically randomized protocol. We can emulate clamped thermalization for the visibles being clamped to this mixed state via classical randomization by modifying step 3 of the regular algorithm as follows:

\begin{enumerate}
\item[3.] Clamped thermalization 
\begin{enumerate}
\item Initialize pulse parameters $\bm{\gamma}^{(n,0)}$ and $\bm{\nu}^{(n,0)}$ as random
\item Prepare visible units as computational basis state corresponding to a data point $\ket{\bm{x}}_v$ uniformly randomly chosen from the dataset $D$.
\item Prepare hidden units in a thermal state of $H_{\tilde{M}}$ of inverse temp $\beta$ by entangling qubits in pairs
\item Apply QAOA circuit via $P$ simulated pulses alternating between partial cost and partial mixer Hamiltonian evolution \[\qquad\qquad \prod_{l=1}^P \exp(-i\nu^{(n,m)}_l H_M)\exp(-i\gamma^{(n,m)}_l H_C)\]
\item Measure all qubits in standard (computational basis) basis
\item Repeat (b)-(e) to estimate cost Hamiltonian expectation value $\braket{\psi_{n,m}|H^{(n)}_{\tilde{C}}|\psi_{n,m}}$  
\item Using classical optimizer, figure out updates $\bm{\gamma}^{(n,m+1)}$ and $\bm{\nu}^{(n,m+1)}$ to minimize $\braket{H^{(n)}_F}$

\item Repeat (b)-(g) until minimum of $\braket{H^{(n)}_F}$ reached; optimal  $\bm{\gamma}^{(n)}$ and $\bm{\nu}^{(n)}$ QAOA parameters are found.
\item Measure \& register expectation values $\braket{Z_jZ_k}$ and $\braket{Z_j}$ for optimal QAOA circuit from (g)
\end{enumerate}
\end{enumerate}

And step 4, once again the averaged expectation value for observables $\overline{\braket{\ldots}}_D$ is replaced by the randomized clamped expectation value from the above process.

An option with this training in batches would be to perform a \textit{quantum stochastic gradient descent}, i.e.~ by clamping to a different \textit{minibatch} (randomized subset of the data \cite{Goodfellow-et-al-2016}) for each epoch (iteration of the weight update).

\section{Data generation}

In this appendix we show how the datasets used to generate the results in figure \ref{fig:KL_noises} and \ref{fig:sampling} were constructed. As we are trying to evaluate how well our Boltzmann machines can find hidden variables in data, we construct data sets which have data encoded in a lower-dimensional hidden subspace. 

For the data in figure \ref{fig:KL_noises}, we sample from $k$ Bernoulli random variables, encoding them into $n$ bits using a classical linear $[n,k]$ code \cite{Gottesman1997}. We then add some independent bit flip noise to each of the $n$ bits. More specifically, the Bernoulli variables were sampled from a distribution where $p(0) =1- \eta$, $p(1) =\eta$, $\eta= 0.6$, the classical linear code was a $[4,2]$ repetition code, and the individual bit flip noise had a flipping probability of $p=0.025$. We evaluated the KL divergence by testing with new encoded data points versus the hidden units. The KL can then be seen as a measure of decoding success, in a sense.

As for the data used for figure \ref{fig:sampling}, we used a hidden mode data similar to \cite{QRBM}. This consists of having so-called hidden Bernoulli modes: consider a set of bit strings $\bm{m}_j \in\{0,1\}^n$, where $j\in\{1,\ldots,k\}$ and the probability of a $n$-bit string $x$ being input into the visibles is
\begin{equation}
P_V(\bm{x})  = \tfrac{1}{k}\textstyle\sum_{j=1}^k p^{n-\lVert \bm{x}-\bm{m}_j\rVert_0}(1-p)^{\lVert \bm{x}-\bm{m}_j\rVert_0}
\end{equation}
where $p\in[0,1]$ and $\lVert \bm{x}-\bm{m}_j\rVert_0$ is the Hamming distance between the bit string $\bm{x}$ and the mode string $\bm{m}_j$. We chose $n=4$ visible units, $k=2$ hidden modes, and $p=0.9$ for the data depicted in figure \ref{fig:sampling}.

\section{Analog vs. Approximate Thermalization} 
\subsection{Quantum Annealing}
It is  worth recalling the traditional, non-circuit based approach, to quantum enhanced Gibbs sampling.  Quantum annealers, such as D-Wave \cite{2016arXiv161104528K}, offer a means to physically implement the thermalization process according to a pre-programmed Hamiltonian. Hence the physical system itself implements \textit{analog} Gibbs sampling. The way annealers achieve thermalization of a cost (target) Hamiltonian, e.g.~a Hamiltonian of the form $H_C\equiv\sum_{j,k} J_{jk} Z_jZ_k + \sum_{j} B_j Z_j ,$
is by starting with the thermal state of a Hamiltonian with a  known an accessible ground state (e.g.~$H_M = \sum_j X_j$)
and allowing open system evolution as they slowly change the Hamiltonian
$H(\tau) = (1-\tau) H_M + \tau H_C,$  $ \tau\in [0,1]. $

Ideally, the state remains in the thermal state of the instantaneous Hamiltonian at all times $e^{-\beta H(\tau)}/\mathcal{Z}_\tau$, $\mathcal{Z}_\tau = \text{tr}(e^{-\beta H(\tau)})$ beginning in the initial Hamiltonian thermal state $ e^{-\beta H_M}/\mathcal{Z}_M = \bigotimes_{j} e^{-\beta X^j}/\mathcal{Z}_j $
where $\mathcal{Z}_M= \prod_j \mathcal{Z}_j,\ \mathcal{Z}_j = \text{sech}(\beta)/2$, and ending in the cost Hamiltonian's thermal state, \begin{equation}
    \tfrac{1}{\mathcal{Z}_C}e^{-\beta H_C } = \tfrac{1}{\mathcal{Z}_C}\sum_j e^{-\beta E_j} \ket{E_j}\!\bra{E_j}
\end{equation}  
where 
$\mathcal{Z}_C= \text{tr}\left(e^{-\beta H_C}\right). $ This should be the case as $H(\tau)$ is swept slowly enough so that the open dynamics allow for thermalization on a time scale smaller than the annealing time. 
\subsection{Quantum Approximate Thermalization}

We start the protocol by preparing the purified thermal state $
 \ket{\psi_0}=\bigotimes_{j=1}^Q \mathcal{Z}_j^{-1/2} \textstyle\sum_{b\in\{0,1\}} e^{(-1)^{1+b}  \beta/2 }\ket{b}_j\ket{b}_{E_j}$
using a set of environment purification registers $E = \bigotimes_{j=1}^Q E_j$, and with $Q$ the total number of qubits. The state with the environment qubits traced out is the thermal state of $H_I$, i.e. $e^{-\beta H_I}/\mathcal{Z}_I$. We can write this thermal state with respect to the initial Hamiltonian as a classical mixture of bit strings 
 \begin{equation}\label{eq:init}
     \rho_I =  \tfrac{1}{\mathcal{Z}_I}\textstyle \sum_{\bm{k}\in\{0,1\}^N} e^{-\beta |\bm{k}|} \ket{\bm{k}}\!\bra{\bm{k}}
 \end{equation} where $\mathcal{Z}_I = \text{sech}^Q(\beta)/2^Q$ and we denote $\ket{\bm{k}} \equiv \bigotimes_{j=1}^Q \ket{k_j}$ and $|\bm{k}| = \lVert\bm{k}\rVert_0$.
 We find that the spectrum of the density operator is given by
$\text{spec}(\rho_M) =  \left\{\lambda_{\bm{k}}\right\}_{\bm{k} } ,$  $\lambda_{\bm{k}}\equiv e^{-\beta |\bm{k}|}2^{-Q}\text{sech}^Q(\beta)  
$
which is a decaying exponential function of the Hamming weights of all $Q$-bit strings.
 
The QAOA can be seen as approximately simulating the adiabatic evolution interpolating from $H_M$ to $H_C$. Thermodynamically, one can see this process as simulating having an initial open system thermal equilibrium with respect to the initial Hamiltonian $H_I$, thus introducing entropy into the system, then evolving the system according to a \textit{closed} quantum system unitary evolution, in a way to (approximately) maintain an equilibrium state. Since QAOA minimizes energy, and our evolution fixes the entropy initially, we can think of the QAOA circuit as minimizing energy at fixed entropy ($S$), which is a way to minimize the final free energy $F_C= \braket{H_C}-\tfrac{1}{\beta }S$. As is well known in standard quantum thermodynamics, minimizing the free energy, brings our system state closer to the thermal state, in terms of relative entropy.

As we apply unitary evolution to this density operator, the spectrum will be necessarily conserved. We can consider the state of minimal energy achievable through unitary evolution as
\begin{equation}\label{eq:perf}
\rho_\text{perf} = \!\!\!\!\sum_{\bm{k} \in\{0,1\}^N}\!\!\! \lambda_{\bm{k}} \ket{E_{\bm{k}}}\!\bra{E_{\bm{k}}} = \tfrac{1}{Z_M}\!\!\!\!\sum_{\bm{k} \in\{0,1\}^N}\!\!\!e^{-\beta |{\bm{k}}|} \ket{E_{\bm{k}}}\!\bra{E_{\bm{k}}}
\end{equation}
where the energy eigenstates are indexed such that higher energies have a higher Hamming weight index $E_{\bm{k}} \leq E_{\bm{j}} \implies |\bm{k}| \leq |\bm{j}|$. 
We can argue the above form is the optimum of energy minimization over all possible unitaries, up to reshuffling of equal eigenvalues for degenerate energy levels---thus it is the minimal energy state achievable through unitary evolution.

Note that even in the idealized (infinite) QAOA limit, we do not obtain an exact thermal state of the target Hamiltonian $H_C$. What we effectively obtained is thus a ``thermal-like'' state, but not quite thermal, since the spectrum is not quite that of the true thermal state of the cost Hamiltonian. The relative entropy of this \textit{pseudo-thermal} state to the actual thermal state of $H_C$ can be computed as:
$D(\rho||\rho_C) = \log(\mathcal{Z}_C) +\tfrac{\beta}{\mathcal{Z}_M}\sum_{\bm{k}} e^{-\beta \bm{k}} E_{\bm{k}} - Q\beta \tanh(\beta)$. For small $\beta$ the relative entropy tends to zero, which is to be expected as this case would reduce to a ground state optimization, and QAOA was originally designed for such problems.



%

\end{document}